\documentclass[%
 reprint,
 superscriptaddress,
 amsmath,amssymb,
 aps,
 prapplied,
]{revtex4-2}

\usepackage{subcaption}

\usepackage{algorithm, algpseudocode}

\usepackage{tikz}
\usepackage{quantikz}
\usepackage{braket}
\usepackage{comment}

\usepackage{color}

\usepackage{graphicx}
\usepackage{dcolumn}
\usepackage{bm}

\usepackage[justification=raggedright,singlelinecheck=false]{caption}

\begin{document}

\title{Quantum State Preparation for Probability Distributions with Reflection Symmetry Using Matrix Product States}

\author{Yuichi Sano}
 \email{sano.yuichi.s98@kyoto-u.jp}
\affiliation{%
 Department of Nuclear Engineering, Kyoto University, Nishikyo-ku, Kyoto 615-8540, Japan.
}%

\author{Ikko Hamamura }
\altaffiliation[Present Address: ]{NVIDIA G.K., Tokyo 107-0052, Japan.}
\affiliation{
IBM Quantum, IBM Japan 19-21 Nihonbashi Hakozaki-cho, Chuo-ku, Tokyo, 103-8510, Japan.
}%

\date{\today}

\begin{abstract}
Quantum circuits for loading probability distributions into quantum states are essential subroutines in quantum algorithms used in physics, finance engineering, and machine learning.
The ability to implement these with high accuracy in low-depth quantum circuits is a critical issue.
We propose a novel quantum state preparation method for probability distribution with reflection symmetry using matrix product states.
By considering reflection symmetry, our method reduces the entanglement of probability distributions and improves the accuracy of approximations by matrix product states. 
As a result, we improved the accuracy by two orders of magnitude over existing methods using matrix product states.
Our approach, characterized by linear scalability with qubit count, is highly advantageous for noisy quantum devices.
Also, our demonstration results reveal that the approximation accuracy in tensor networks depends heavily on the bond dimension, with minimal reliance on the number of qubits.
Our method is demonstrated for a normal distribution encoded into 10 and 20 qubits on a real quantum processor.

\end{abstract}

\maketitle

\section{Introduction}
In various quantum algorithms such as the Monte Carlo method by quantum computer\cite{Rebentrost18,Orus2019,Egger20}, quantum machine learning\cite{Biamonte2017,Havlicek2019,Liu2021}, and simulations for physics\cite{Preskill2012,li2023potential}, quantum state preparation constitutes a vital subroutine.
The efficiency of quantum state preparation becomes vital in determining whether these algorithms can achieve superiority over classical algorithms.

It is known that the preparation of quantum states for arbitrary functions without error requires an exponential depth in quantum circuits\cite{Grover2002,zalka1998simulating}.
In recent years, several methods have been proposed for state preparation using lower depth quantum circuits by tolerating a degree of error.
In quantum machine learning, a proposed method involves creating quantum circuits to generate target probability distributions, utilizing quantum Generative Adversarial Networks with parameterized quantum circuits as Generators\cite{Zoufal2019,SKNA2023}.
Methods utilizing approximations via Fourier transforms have also been proposed\cite{markov2022generalized,Moosa_2024}.
Furthermore, approaches utilizing tensor networks have also been actively researched in recent times\cite{Holmes2020,Ran2020,iaconis2023quantum,Melnikov2023}.

We propose a new state preparation method utilizing the reflection symmetry of probability distributions and tensor networks\cite{Ran2020,iaconis2023quantum}.
We note that functions with less entanglement can be approximated more accurately when using matrix product states.
Specifically, in the normal distribution, the portion to the left of the mean monotonically increases and possesses a structure with less entanglement.
Therefore, using matrix product states, we could convert the left half of the normal distribution into quantum circuits with higher accuracy.
Subsequently, we add a quantum circuit that ``duplicates'' by the Hadamard gate and inverts this half by CNOT gates.
This quantum circuit has enabled us to prepare the entire normal distribution from the quantum circuit that prepares the left half of the normal distribution.
Our method achieves a two-order-of-magnitude improvement in normal distribution accuracy compared to methods using existing matrix product states.
In addition, we have also confirmed that accuracy improvements can be similarly achieved with distributions with reflection symmetry other than the normal distribution.
Our method features a quantum circuit in which most parts consist of gates between nearest-neighbor qubits. 
The circuit exhibits an extremely low circuit depth that scales linearly with the number of qubits. 
This makes it a beneficial property for noisy quantum devices.
We showed that when using tensor networks, the approximation accuracy heavily depends on the bond dimension of the tensor network and is mainly independent of the number of qubits.
Finally, we demonstrated that we lord the normal distribution into 10 and 20 qubits on a real quantum processor using our designed quantum circuits, achieving high fidelity.

\section{Tensor Networks to Quantum Circuits}
In this section, we introduce a method for converting Matrix Product States (MPS), a type of tensor network, into quantum circuits\cite{Ran2020,iaconis2023quantum,PhysRevLett.95.110503}.

Any quantum state of finite dimension can be exactly represented using MPS.
If we consider a quantum state as a single tensor and perform singular value decomposition (SVD), we can quickly obtain a representation of the quantum state using MPS.
However, the bond dimension between the sites of the MPS becomes very large.
Generally, a method for exactly converting such MPS into a quantum circuit has yet to be discovered.

\begin{figure}[t]
\begin{minipage}[b]{0.7\linewidth}
\centering
\includegraphics[width=\linewidth]{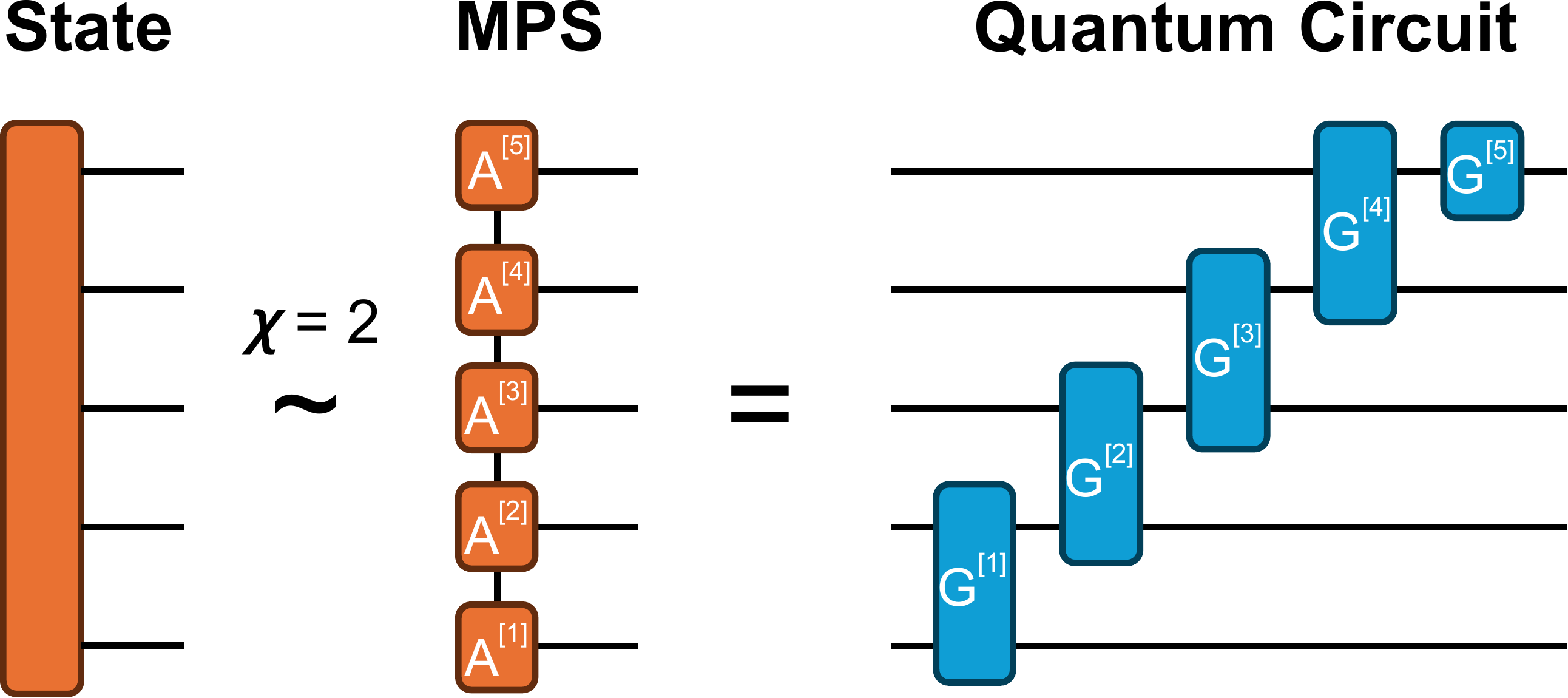}
\subcaption{}
\label{fig:TNtoQC_a}
\end{minipage}
\begin{minipage}[b]{0.9\linewidth}
\centering
\includegraphics[trim={5mm 0 0 0},clip,width=\linewidth]{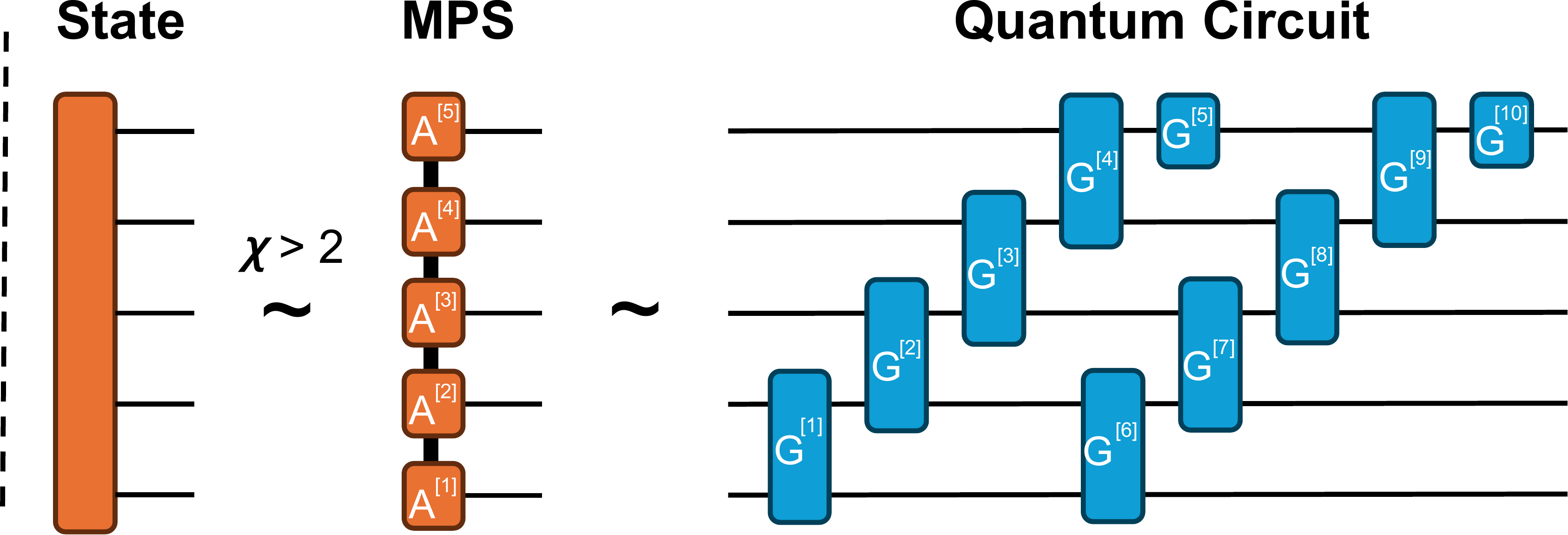}
\subcaption{}
\label{fig:TNtoQC_b}
\end{minipage}
\caption{Method for converting a quantum state (five qubits) into a quantum circuit. (a) If the bond dimension $\chi = 2$, the conversion to MPS is performed approximately, while the conversion to a quantum circuit is executed precisely. (b) If the bond dimension $\chi > 2$, the conversion to MPS and the conversion to a quantum circuit are performed approximately.}
\label{fig:TNtoQC}
\end{figure}

MPS consisting of $n$ qubits, as shown in Figure \ref{fig:TNtoQC}(a), can be written as follows:
\begin{equation}
    \ket{\Psi}=\sum_{a_1, \cdots, a_{n-1}} \sum_{s_1, \cdots, s_n} A_{s_1, a_1}^{[1]} A_{s_2, a_1 a_2}^{[2]} \cdots A_{s_n, a_{n-1}}^{[n]} \ket{s_1,\cdots, s_n},
\end{equation}
where  $s_i \in \{0, 1\}$ represents a physical index, and $a_i \in \{0, \ldots, \chi - 1\}$ is a virtual index.
When converting a quantum state to MPS, the bond dimension $\chi$ can be restricted to any values by limiting the count of singular values in each SVD\cite{Schollw_ck_2011,PhysRevLett.69.2863}.
Furthermore, in this paper, it is assumed that the MPS satisfies the following left canonical form:
\begin{align}
& \sum_{s_1 a_1} A_{s_1, a_1}^{[1]} A_{s_1, a_1}^{[1] *}=1, \\
& \sum_{s_i a_i} A_{s_i, a_{i-1} a_i}^{[i]} A_{s_i, a_{i-1}^{\prime} a_i}^{[i] *}=I_{a_{i-1} a_{i-1}^{\prime}}, \\
& \sum_{s_n} A_{s_n, a_{n-1}}^{[n]} A_{s_n, a_{n-1}^{\prime}}^{[n] *}=I_{a_{n-1} a_{n-1}^{\prime}},
\end{align}
where $1<i<n$ in Equation (3).
From these equations, it can be understood that each $A^{[i]}$ is an isometry.

First, we describe the method when the bond dimension is limited to $\chi = 2$.
Our objective is to obtain a unitary operator $U_{{\rm MPD}}$, referred to as a Matrix Product Disentangler (MPD), which operates as follows:
\begin{equation}
    U_{{\rm MPD}} \ket{\Psi} = \ket{0}^{\otimes n}.
\end{equation}
We construct a single-qubit gate $G^{[n]}$ from the last tensor $A^{[n]}$ as follows:
\begin{equation}
    G^{[n]}=A^{[n]}.
\end{equation}
Then, for $i$, such as $1<i<n$, we determine the gate elements as follows:
\begin{equation}
    G_{0 k l m}^{[i]}=A_{k l m}^{[i]},
\end{equation}
where $G_{jklm}$ represents the matrix element of the gate $G^{[i]}$, with $j$ denoting the computational basis state of the reflection qubit, $k$ corresponding to the input qubit from the previous gate, and $l$ and $m$ representing the computational basis states of the output qubits.
Recalling that $A^{[i]}$ is an isometry, we can readily extend $G^{[i]}$ to be a unitary gate.
Then, we determine the elements of the gate $G^{[1]}$ from the first tensor $A^{[1]}$ as follows:
\begin{equation}
    G_{00 l m}^{[1]}=A_{l m}^{[1]},
\end{equation}
where $G_{jklm}$ represents the element of the gate $G^{[1]}$, and $j$ and $k$ represents the computational basis state of the reflection qubit.
Similarly, we can extend $G^{[1]}$ as a unitary gate.
Using these gates $G^{[i]}$, the $U_{{\rm MPD}}$ can be constructed, as shown in Figure \ref{fig:TNtoQC}(a), as follows:
\begin{equation}
    U_{{\rm MPD}}^{\dagger} = \prod_{i=1}^n G^{[i]},
\end{equation}
where $G^{[i]}$ ($1 \leq i<n$) acts on the $i$-th and $i$+1-th qubits, and $G^{[n]}$ acts on the $n$-th qubit.

\begin{figure}[t]
    \centering
    \includegraphics[width=\columnwidth]{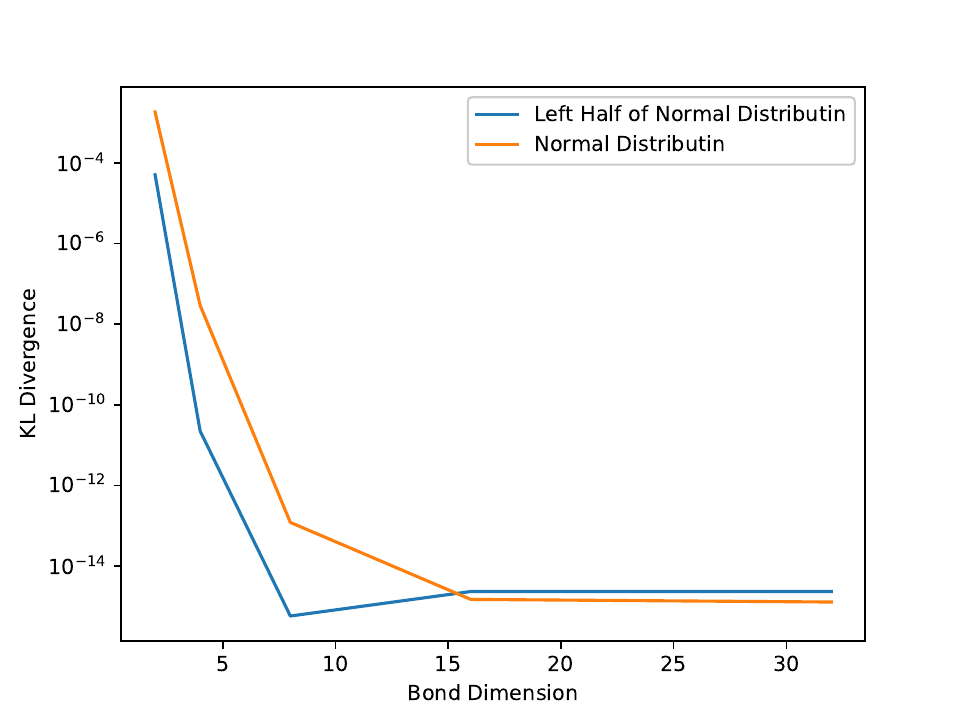}
    \caption{The Kullback-Leibler divergence between the normal distribution $\mathcal{N}(\mu=0, \sigma^2 = 0.01)$ $(\mathrm{min}:-0.5, \mathrm{max}:0.5)$ (or the left half of normal distribution) and the probability distributions represented by each MPS (10 sites) with limited bond dimension.}
    \label{fig:mps_bonddim}
\end{figure}

Next, we describe the method when the bond dimension is limited to $\chi > 2$.
The following approach is based on reducing the bond dimension by using $U_{{\rm MPD}}$.
For instance, we consider the state $\ket{\Psi^{(4)}}$ when $\chi = 4$.
We approximate $\chi = 2$, and construct $U_{{\rm MPD}}^{(2)}$ using the aforementioned method.
Then, using this $U_{{\rm MPD}}^{(2)}$, we disentangle $\ket{\Psi^{(4)}}$.
In other words, we consider $U_{{\rm MPD}}^{(2)} \ket{\Psi^{(4)}}$ to be MPS with approximately $\chi = 2$ and approximate $U_{{\rm MPD}}^{(2)} \ket{\Psi^{(4)}}$ again to $\chi = 2$ to construct $U_{{\rm MPD}}^{(4)}$.
Therefore, it can be considered that $U_{{\rm MPD}}$ of the state $\ket{\Psi^{(4)}}$ can be approximated as follows:
\begin{equation}
    U_{{\rm MPD}} \sim U_{{\rm MPD}}^{(2)} \cdot U_{{\rm MPD}}^{(4)}.
\end{equation}
Using a similar approach, in the case of $\chi=2^L$, the $U_{{\rm MPD}}$ can be constructed approximately as follows:
\begin{equation}
    U_{{\rm MPD}} \sim \prod_{l=1}^L U_{{\rm MPD}}^{(2^l)}.
\end{equation}

\section{Quantum Circuit for Normal Distribution Utilizing Symmetry}
In this section, we show that we can prepare quantum circuits with higher accuracy by considering the reflection symmetry of normal distributions.

\subsection{Entanglement and Symmetry}

\begin{figure}[t]

\begin{minipage}[b]{0.49\linewidth}
\centering
\resizebox{0.95\linewidth}{!}{%
\begin{quantikz}[row sep = 0.7em, thin lines]
  \lstick{$\ket{0}$} & \qw    \slice{(1)}                  & \gate{H}\slice{(2)} & \ctrl{1} & \ctrl{2} & \ctrl{3} & \ctrl{4}\slice{(3)} & \qw\\
  \lstick{$\ket{0}$} & \gate[wires=4][0.5cm]{U_{{\rm MPD}}} & \qw      & \targ{}  & \qw      & \qw      & \qw      & \qw\\
  \lstick{$\ket{0}$} & \qw                                  & \qw      & \qw      & \targ{}  & \qw      & \qw      & \qw\\ 
  \lstick{$\ket{0}$} & \qw                                  & \qw      & \qw      & \qw      & \targ{}  & \qw      & \qw\\
  \lstick{$\ket{0}$} & \qw                                  & \qw      & \qw      & \qw      & \qw      & \targ{}  & \qw
\end{quantikz}
}
\subcaption{}
\label{fig:QC_a}
\end{minipage}
\begin{minipage}[b]{0.49\columnwidth}
\centering
\includegraphics[width=\columnwidth]{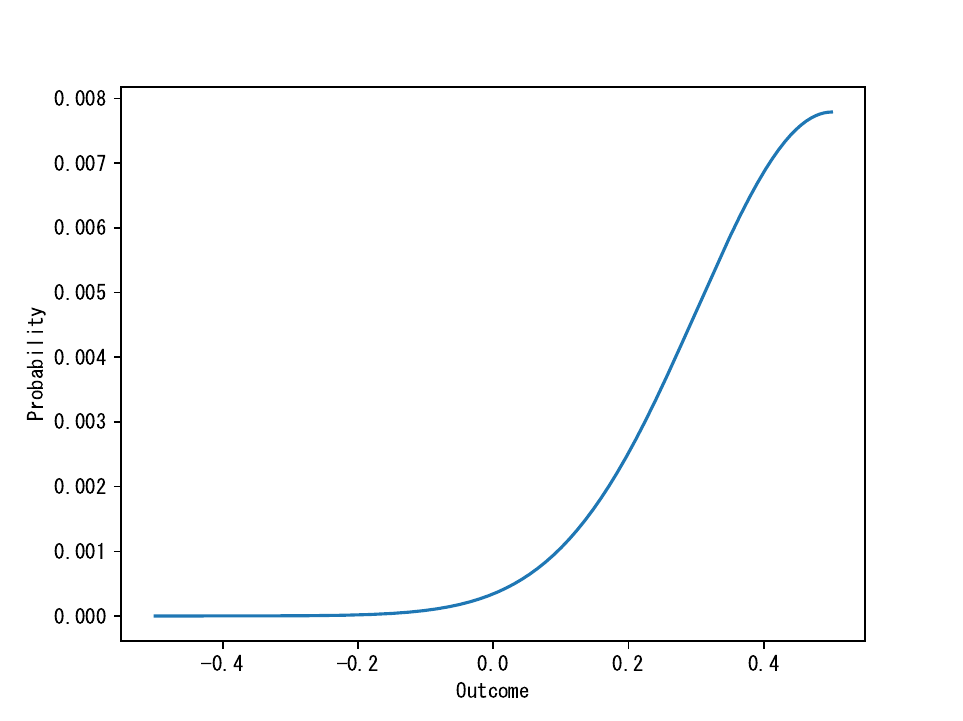}
\subcaption{}
\label{fig:QC_b}
\end{minipage}
\begin{minipage}[b]{0.49\columnwidth}
\centering
\includegraphics[width=\columnwidth]{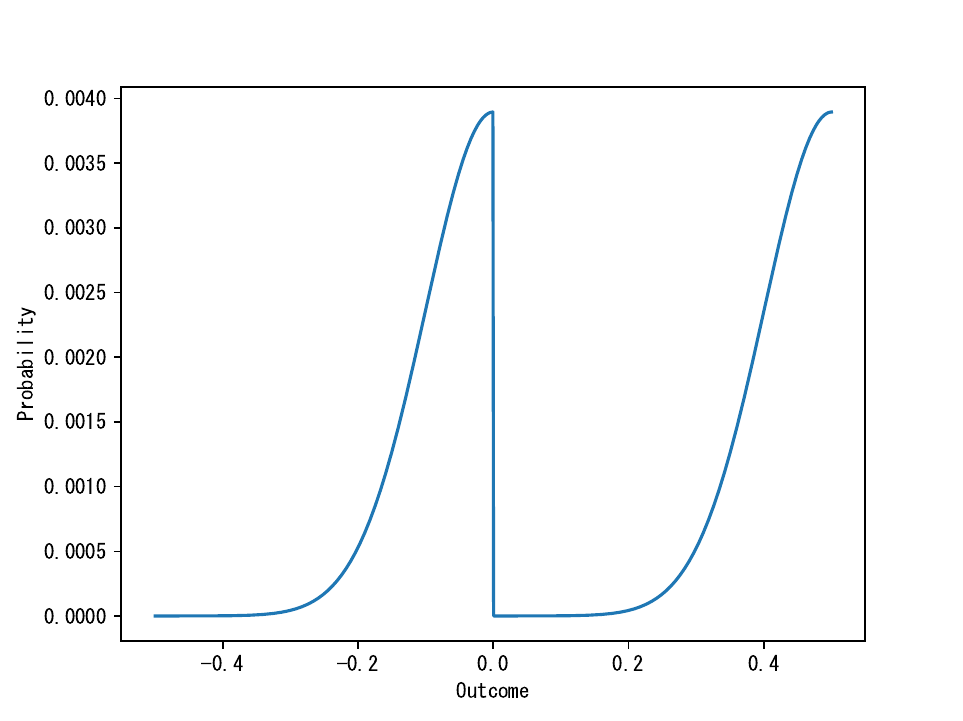}
\subcaption{}
\label{fig:QC_c}
\end{minipage}
\begin{minipage}[b]{0.49\columnwidth}
\centering
\includegraphics[width=\columnwidth]{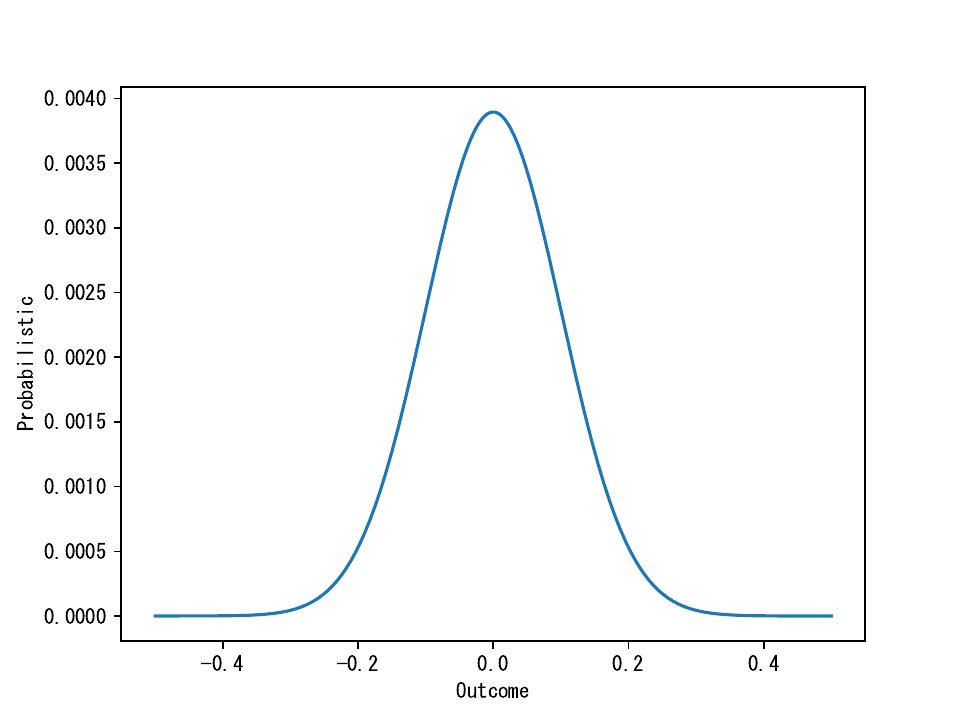}
\subcaption{}
\label{fig:QC_d}
\end{minipage}
    \caption{Quantum circuit considering the reflection symmetry of the normal distribution (five qubits). (a) If a quantum circuit $U_{{\rm MPD}}$ exists that creates the left half of a normal distribution, it can be extended to the right half. 
    Let us call the highest qubit that is used for this reflection the reflected qubit.
    (b) At the point of (1), a quantum state with the left half of the normal distribution as its probability amplitude is created by the quantum circuit $U_{{\rm MPD}}$. (c) At the point of (2), the left half of the normal distribution is duplicated by applying the Hadamard gate to a reflection qubit. (d) At the point of (3), by applying CNOT gates with the reflection qubit as the control qubit to other qubits, the left half of the right-side normal distribution is transformed into the right half.}
    \label{fig:quantumcircuit}
\end{figure}

In general, it is difficult to accurately reproduce an arbitrary probability distribution using an MPS with a low bond dimension, and we confirmed this in the case of the normal distribution in Figure \ref{fig:mps_bonddim}.
Lower bond dimensions cause larger errors because the bond dimension is related to the entanglement of the quantum state.
This implies that probability distributions with less entanglement will likely have smaller errors than those with larger entanglement.

We consider that segmenting the probability distribution to reduce entanglement could allow us to prepare the distribution with greater precision.
For example, while the entanglement measure of the normal distribution $\mathcal{N}(\mu=0, \sigma^2 = 0.01)$ is $0.237 \times 10^{-2}$, the entanglement measure of the left half of the normal distribution is $8.31 \times 10^{-6}$.
Therefore, it is expected that the half-normal distribution can be approximated with low entanglement, and indeed, as shown in Figure \ref{fig:mps_bonddim}, we found that the approximation accuracy for the half-normal distribution saturates at a lower bond dimension.
Remark that for the entanglement measure, we used the following measure of multiple qubit entanglement proposed in previous research \cite{Meyer2002}:
\begin{equation}
    Q(\psi)=\frac{4}{n} \sum_{j=1}^n D\left(\imath_j(0) \psi, \imath_j(1) \psi\right),
\end{equation}
where $D(\cdot,\cdot)$ represents
\begin{equation}
D(u, v)=\sum_{x<y}\left|u_x v_y-u_y v_x\right|^2,
\end{equation}
and $\imath_j$ represents
\begin{equation}
    \imath_j(b)\ket{b_1 \ldots b_n}=\delta_{b b_j}\ket{b_1 \ldots \widehat{b_j} \ldots b_n}.
\end{equation}

We have discovered that we can reproduce the normal distribution with high precision if we only consider half of it, but we need to prepare the remaining right half as well. 
Here, the reflection symmetry of the normal distribution becomes significant.
We propose a quantum circuit that can ``copy'' the right half of the normal distribution from the left half, as shown in Figure \ref{fig:quantumcircuit}.
Therefore, we can create a high-precision normal distribution from its high-accuracy left half.

The quantum circuit in Figure 3 can, in a sense, be regarded as a form of $U_{MPD}$ that successfully extracts the entanglement structure of a probability distribution with reflection symmetry.
This can be seen as an example of incorporating prior knowledge to extract the entanglement structure of a quantum state—something that could not be achieved by constructing $U_{MPD}$ solely through MPS and the Disentangler.

\subsection{Result}
\begin{figure}[tb]
\begin{minipage}[b]{0.8\columnwidth}
\centering
\includegraphics[width=\columnwidth]{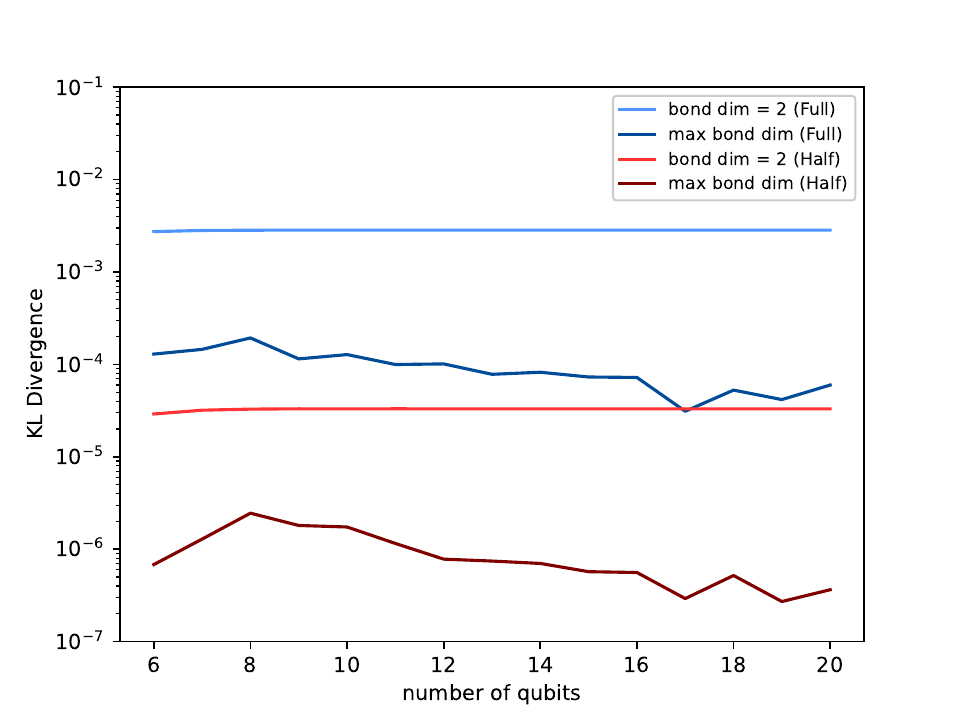}
\subcaption{}
\label{fig:result_1}
\end{minipage}
\begin{minipage}[b]{0.8\columnwidth}
\centering
\includegraphics[width=\columnwidth]{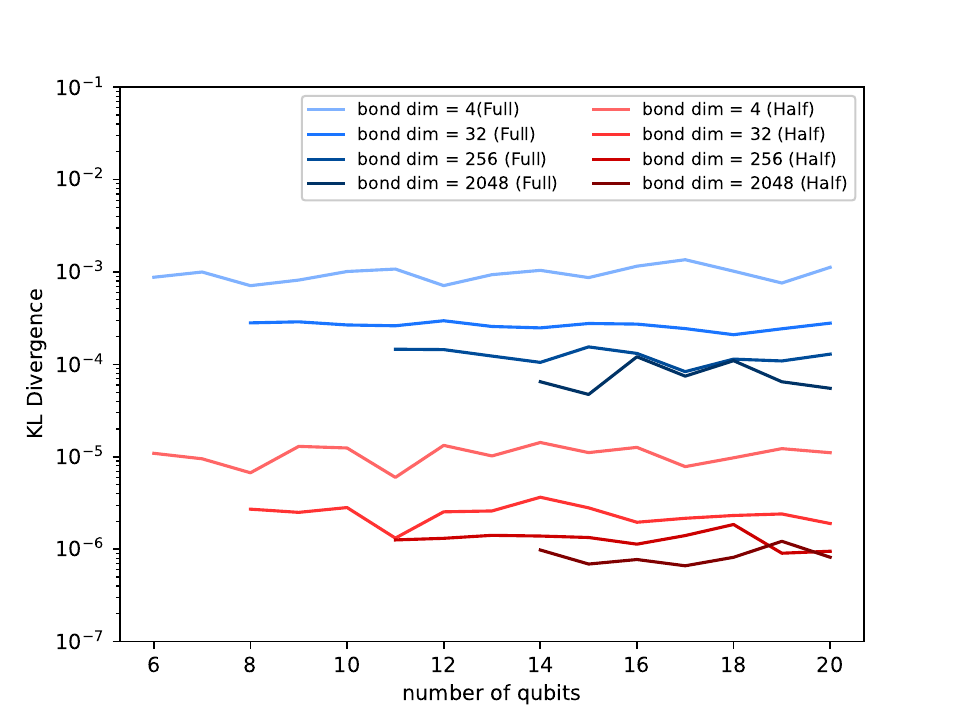}
\subcaption{}
\label{fig:result_2}
\end{minipage}
    \caption{Comparison of the KL divergence between the existing and our methods. (a) The KL divergence is when the bond dimension is $\chi = 2$ and maximum ($\chi = 2^{n-1}$). Our method demonstrates an order of magnitude accuracy better than the existing method. (b) When converting from tensor networks to quantum circuits, the KL divergence depends on the bond dimension and shows largely independence on the number of qubits.}
    \label{fig:result}
\end{figure}

We converted the normal distribution $\mathcal{N}(\mu=0, \sigma^2 = 0.01)(\mathrm{min}:-0.5, \mathrm{max}:0.5)$ into a quantum circuit using our method and then calculated the KL divergence between the quantum state generated by the quantum circuit and the original probability distribution.
We constructed MPS using Tensor Networks \cite{Roberts2019} and executed quantum circuits derived from these MPS using Qiskit \cite{Qiskit2023}.

The results are presented in Figure \ref{fig:result}.
Figure 4(a) shows that our approach improved the KL divergence by approximately $10^{-2}$.
Additionally, by considering the bond dimension, we found it feasible to approximate the probability distribution with the KL divergence of up to the order of $10^{-7}$.
Figure 4(b) presents the KL divergence when the bond dimension is constant. 
It can be observed that the KL divergence is generally independent of the number of qubits and depends primarily on the bond dimension.
This suggests that, for approximating the normal distribution, the procedure described in Section 2 can be repeated 11 times to achieve the KL divergence as $\mathcal{O}(10^{-7})$, regardless of the qubit count.

Finally, we demonstrated our methods with a quantum processor \textit{ibm\_torino} using 10 and 20 qubits.
We converted and learned the normal distribution $\mathcal{N}(\mu=0, \sigma^2 = 1)(\mathrm{min}:-4\sqrt{3}, \mathrm{max}:4\sqrt{3})$ and measured the final state with 100,000 shots for 10-qubit demonstration and 3,000,000 shots for 20-qubits demonstration.
Several error mitigation methods were utilized by the sampler in qiskit runtime primitives.
M3 readout error mitigation~\cite{Nation2021} was performed for the 10-qubit demonstration but not for the 20-qubit one because it timed out.
Dynamical decoupling~\cite{Hahn1950, Ezzell2023} was also enabled.

\begin{figure}[t]
    \begin{minipage}[b]{0.8\columnwidth}
    \centering
    \includegraphics[width=\columnwidth]{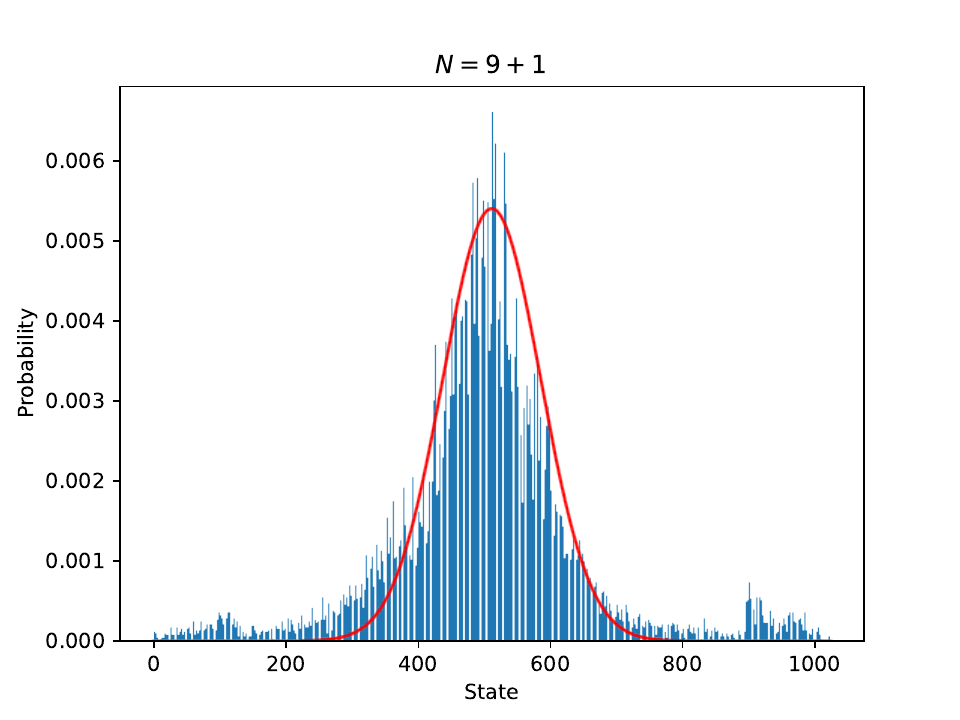}
    \subcaption{}
    \end{minipage}
    \begin{minipage}[b]{0.8\columnwidth}
    \centering
    \includegraphics[width=\columnwidth]{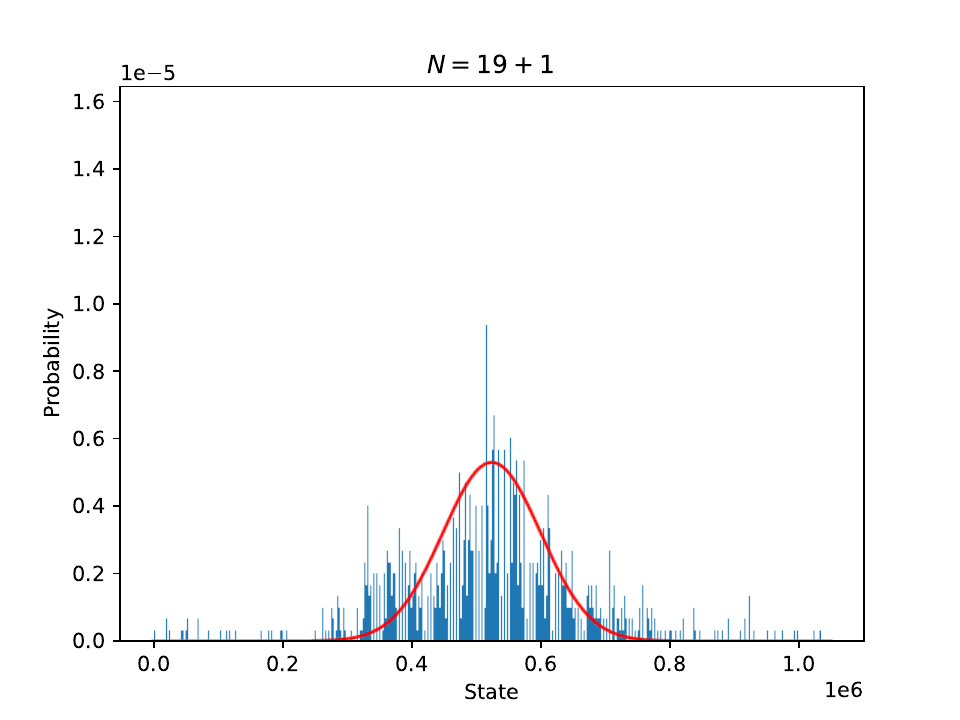}
    \subcaption{}
    \end{minipage}
    \caption{
    Sampled normal distributions by \textit{ibm\_torino}.
    The qubits are 10 qubits for (a) and 20 qubits for (b).
    The number of samples is 100,000 for 10 qubits and 3,000,000 for 20 qubits.
    The fidelity for the probability distributions is 0.879 for 10 qubits and 0.795 for 20 qubits.
}

    \label{fig:real-experiment}
\end{figure}

The results are shown in Fig.~\ref{fig:real-experiment}.
Both histograms fit the normal distribution well.
Fidelities were 0.879 for 10 qubits and 0.795 for 20 qubits.

\subsection{Other Functions}
\begin{figure}[t]
\begin{minipage}[b]{0.8\columnwidth}
\centering
    \centering
    \includegraphics[width=\columnwidth]{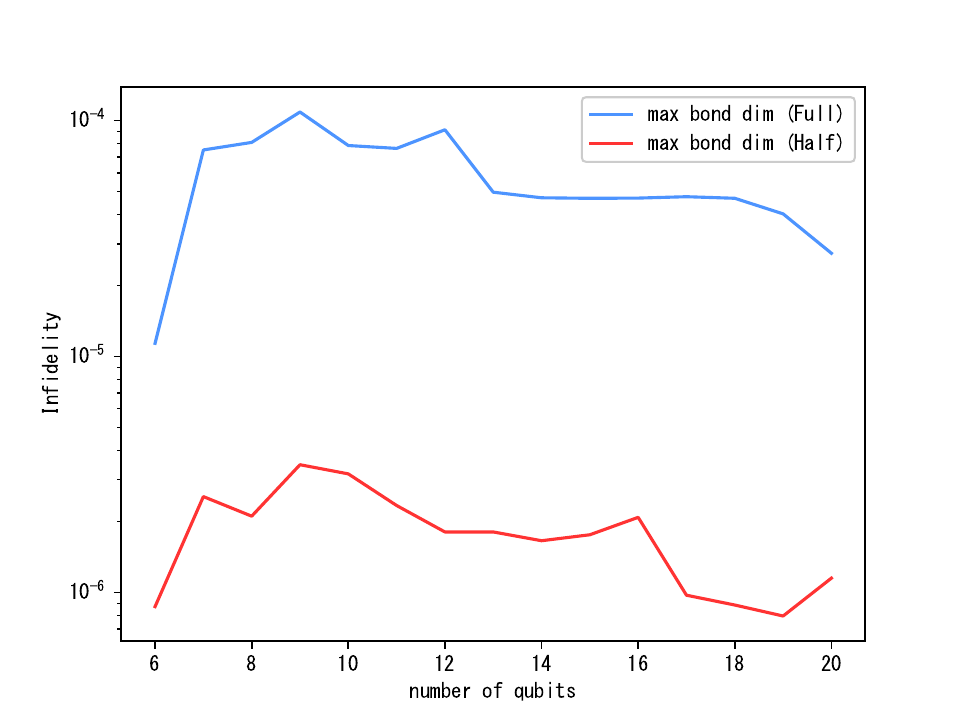}
    \subcaption{}
\label{fig:lorentz}
\end{minipage}
\begin{minipage}[b]{0.8\columnwidth}
\centering
    \centering
    \includegraphics[width=\columnwidth]{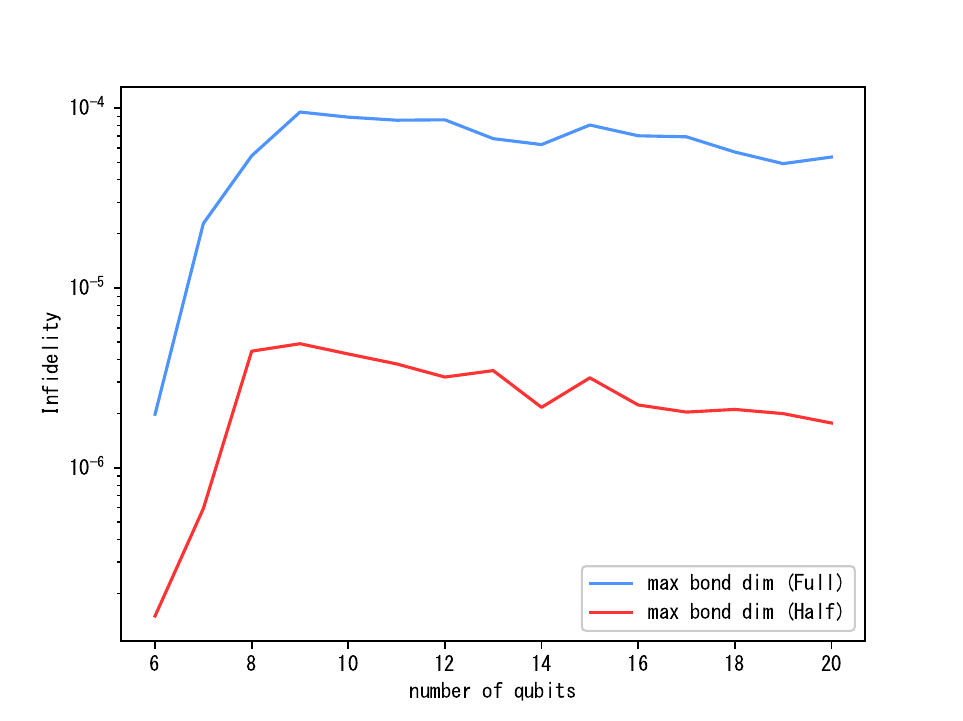}
    \subcaption{}
\label{fig:t_dist}
\end{minipage}
    \caption{Our methodology applied to (a) the Lorentzian function(min: -5, max: 5), and (b) the Student's t-distribution($\nu$ = 2, min: -10, max: 10).The max bond dimension refers to $\chi = 2^{n-1}$.}
    \label{fig:other_func}
\end{figure}

Our method can also be applied to other probability distributions and functions with symmetry. 
For instance, as shown in Figure \ref{fig:other_func}, similar improvements in precision can be achieved with distributions such as the Lorentzian function and the Student's t-distribution.

\section{Conclusion}
We proposed a method that utilizes reflection symmetry to achieve higher precision in preparing the normal distribution when converting from tensor networks to quantum circuits.
The depth of this quantum circuit scales linearly with the number of qubits, even when considering the maximum bond dimension.
Our method improved the accuracy (KL divergence) by two orders of magnitude compared to existing methods using tensor networks. 
We showed how bond dimension and the number of qubits influence the precision when preparing the normal distribution using tensor networks.
Specifically, we showed that increasing the bond dimension improves accuracy, and the number of qubits has a negligible impact on accuracy.
Furthermore, we confirmed that our method improves accuracy even in distributions with reflection symmetry, in addition to the normal distribution.
Finally, we implemented our quantum circuit that loads the normal distribution on a real quantum processor.
We performed demonstrations using 10- and 20-qubit quantum circuits, achieving high fidelity in each case.

\begin{table*}
 \caption{Comparison of various state preparation methods for Normal distributions. $n$ refers to the number of qubits, $p$ to the number of parameters, $m$ to the approximation parameter in the FSL\cite{Moosa_2024}, and, $\chi$ represent the bond dimension. For classical optimization, a circle ($\bigcirc$) indicates necessity, while a cross ($\times$) denotes non-necessity.}
 \label{table:comparison}

\begin{ruledtabular}
\begin{tabular}{lcccc}
 \hline
 & Circuit & Gate &  & Classical \\ 
 Methods & Depths & Counts & Accuracy & Optimization \\ \hline \hline
 Our method ($\chi = 2$) & $\mathcal{O}(n)$ & $\mathcal{O}(n)$ & $\mathcal{O}(10^{-5})$ & $\times$ \\
 Our method ($\chi > 2$) & $\mathcal{O}(\log{\chi}+n)$ & $\mathcal{O}(\log{\chi}+n)$ & $\mathcal{O}(10^{-7})$ &  $\times$  \\
 Tensor Network ($\chi = 2$)\cite{Ran2020,iaconis2023quantum} &  $\mathcal{O}(n)$ & $\mathcal{O}(n)$  & $\mathcal{O}(10^{-3})$ &  $\times$   \\
 Tensor Network ($\chi > 2$)\cite{Ran2020,iaconis2023quantum} &  $\mathcal{O}(\log{\chi}+n)$ & $\mathcal{O}(\log{\chi}+n)$  & $\mathcal{O}(10^{-5})$ &  $\times$   \\
 GR State Preparation\cite{Grover2002} & $\mathcal{O}(2^n)$ & $\mathcal{O}(2^n)$ & exact &  $\times$  \\
 Kitaev-Webb\cite{Kitaev2008} & $\mathcal{O}(poly(n))$ & $\mathcal{O}(poly(n))$ & any & $\times$ \\
 FSL\cite{Moosa_2024} & $\mathcal{O}(n^2,2^m)$ & $\mathcal{O}(n^2,2^m)$ & any & $\times$   \\
 Quantum GAN\cite{Zoufal2019,SKNA2023} & $\mathcal{O}(p)$ & $\mathcal{O}(np)$ & $\mathcal{O}(10^{-3})$ &  $\bigcirc$ \\
 TN + Machine Learning\cite{Melnikov2023} & $\mathcal{O}(p)$ & $\mathcal{O}(np)$ & $\mathcal{O}(10^{-4})$ & $\bigcirc$  \\
 \hline
\end{tabular}
\end{ruledtabular}
\end{table*}

\subsection{Comparison with Prior Works}
Table \ref{table:comparison} compares our method and prior methods.

We provide a rough estimate of the circuit depth and gate count for our method. 
It is known that the $G^{[i]}$ gates, created using tensor networks, can be constructed from two CNOT gates and several single-qubit gates \cite{iaconis2023quantum}.
Furthermore, we demonstrated that considering a bond dimension of $\chi = 2048$, which corresponds to $L=11$, allows us to achieve the accuracy on $\mathcal{O}(10^{-7})$.
Then, the quantum circuit $U_{{\rm MPD}}$ has CNOT gate depth of $2(n-2)$ when the bond dimension is $\chi = 2$, and has CNOT gate depth of $2{(n-2)+(L-1)}$ when the bond dimension is $\chi = 2^{L}$.
For instance, if $n=20$ and $L=11$, the CNOT circuit depth would be 60.
Our method requires not only the $U_{{\rm MPD}}$ but also CNOT gates from a reflection qubit, as is shown between points (2) and (3) in Figure \ref{fig:quantumcircuit}(a).
Therefore, the total CNOT circuit depth for the entire setup becomes either $2(n-2)+n-1$ when $\chi = 2$, and $2{(n-2)+L}+n-1$ when $\chi = 2^L$.

When compared to existing methods using tensor networks\cite{Ran2020,iaconis2023quantum}, the significant difference with our approach lies in the accuracy and the CNOT gates between (2) and (3) in Figure \ref{fig:quantumcircuit}(a).
Regarding accuracy, it is as described above.
The CNOT gate acts only between nearest-neighbor qubits in existing tensor network methods. 
This characteristic in quantum computers with a linear topology means that SWAP gates are not required.
This characteristic, which saves SWAP gates, benefits noisy quantum devices without error correction capabilities.
In our method, as shown before point (1) in Figure \ref{fig:quantumcircuit}(a), the CNOT gate is applied only between nearest-neighbor qubits. 
However, the CNOT gates between points (2) and (3) in Figure \ref{fig:quantumcircuit}(a) act between the reflection qubit and all qubits, necessitating SWAP gates. 
In this respect, our method is at a disadvantage compared to existing methods.

The Grover--Rudolph (GR) state preparation \cite{Grover2002} can prepare exact quantum states, but the number of gates and the circuit depths required increases exponentially with the number of qubits. 
The Kitaev--Webb method \cite{Kitaev2008} allows for similarly accurate state preparation with polynomial circuit depth. 
However, it has been pointed out that for fewer than 15 qubits, it requires a greater circuit depth than the GR method \cite{bauer2021practical}.
Additionally, in the Fourier Series Loader (FSL) method \cite{Moosa_2024} using an m-qubit Fourier transform, the GR state preparation is employed to lord the Fourier coefficients, necessitating a circuit depth of $2^m$. 
To prepare quantum states accurately using the FSL method, a moderately large size for m is necessary (for example, around $m=6$, or more). 
Consequently, this necessitates a significantly large CNOT circuit depth in practice.
Moreover, the inverse quantum Fourier transform, which utilizes approximately $\mathcal{O}(n^2)$ two-qubit gates, also requires a considerably sizeable quantum circuit.
A significant disadvantage of these methods is the requirement to use CNOT gates between all qubits, which necessitates many SWAP gates.
Since implementing a logical SWAP gate requires three CNOT gates, these methods necessitate a more significant number of CNOT gates for implementation.
This requirement presents a challenge for executing noisy quantum devices requiring logical SWAP gates.
In contrast, while our method is less accurate, it excels over these methods in terms of low CNOT circuit depth and fewer SWAP gate requirements.

Finally, we compare our approach with methods that use machine learning \cite{Zoufal2019,SKNA2023,Melnikov2023}. 
These methods' advantage is their ability to determine circuit depth and gate count based on the number of parameters. 
If accurate approximations can be achieved with fewer parameters, these methods can implement lower depth quantum circuits compared to our method and other existing methods that depend on the number of qubits.
However, finding a suitable arrangement of parameters and the parameters themselves is challenging in practice. 
These studies have been limited to discovering quantum circuits with an accuracy on the order of $\mathcal{O}(10^{-4})$.
The fact that the CNOT gates in the quantum circuit act only between nearest-neighbor qubits is beneficial.

In summary, our method offers better precision than approaches using tensor networks or machine learning and maintains a quantum circuit depth comparable to these methods. 
While it is less precise than GR state preparation methods, our method significantly benefits from a much lower depth quantum circuit. 
Therefore, our method is compelling for noisy quantum devices with lower CNOT gate fidelity~\cite{Arute2019,Bharti2022}.

\subsection{Application}
Our method efficiently prepares reflection-symmetric probability distributions—such as the normal distribution, Student’s \(t\)-distribution, and Lorentzian function—which frequently serve as initial states in quantum algorithms across finance and physics.

\subsubsection{Finance}
In financial models, the normal distribution is often used to describe asset returns, portfolio risk, and market uncertainty, particularly in models based on Brownian motion. The t-distribution is also commonly employed due to its heavier tails compared to the normal distribution. Quantum algorithms for option pricing, portfolio optimization, and risk estimation—such as those based on amplitude estimation or Monte Carlo methods—frequently require the initialization of quantum states reflecting such distributions~\cite{finance1, finance2, finance-t}.

\subsubsection{Simulation of Physical Systems}
It is known that, when performing physical simulations of quantum field theory on a quantum computer, one uses a normal distribution as the initial state~\cite{QFT1, QFT2}.
In fluid-dynamics simulations using the Quantum Lattice Boltzmann Method, a normal distribution is also used as the initial state~\cite{fluid_dynamics, fluid_dynamics2}.

\subsubsection{Subroutine for State Preparation}
Our method can also be employed as a subroutine for state preparation. Specifically, techniques have been proposed that use the Lorentz function to approximate various target distributions~\cite{sub-ru}; among these, one can replace the Lorentz function preparation with the Lorentz function generated by our method.

\subsection{Future Work}
We have proposed a method for generating higher-accuracy quantum circuits from matrix product states by utilizing the reflection symmetry of the normal distribution and other functions. 
In this work, we focused on partially monotonically increasing functions, but extending this approach to other types of functions is a task for the future. 
Additionally, the precision of our method is contingent upon the accuracy of the conversion from tensor networks to quantum circuits. 
Hence, proposing methods for more precise conversion is a significant challenge. 
As part of this challenge, it would be intriguing to investigate whether incorporating machine learning~\cite{Zoufal2019,SKNA2023,Melnikov2023,Dborin_2022,Huang2024} into our current method could further enhance accuracy.

\section*{Acknowledgment}
YS would like to thank Takayuki Miyadera for the many helpful comments.
YS would like to thank Hidetaka Manabe for the many helpful discussions about the tensor network.
YS acknowledges the IBM Quantum Researchers Program to access quantum computers.
This work was supported by JSPS KAKENHI Grant Numbers JP23KJ1178.

\nocite{*}

\bibliography{ref}

\end{document}